\documentclass[twocolumn,superscriptaddress,aps,pra,preprintnumbers]{revtex4-1}

\usepackage[english]{babel}
\usepackage{graphicx}
\usepackage{bm}
\usepackage{color}
\usepackage{epstopdf}
\usepackage{amsmath}
\usepackage{amssymb}

\usepackage[urlcolor=blue,colorlinks=true,citecolor=blue,linkcolor=blue,pdfstartview={FitH},bookmarks=false]{hyperref}

\newcommand{\Tr}{\text{Tr}}
\newcommand{\Ln}{\text{ln}}
\graphicspath{{fig/}{./fig/}}

\begin{document}

\title{Lifshitz transitions induced by magnetic field}

\author{Andrzej Ptok}
\email[E-mail me at: ]{aptok@mmj.pl}
\affiliation{Institute of Nuclear Physics, Polish Academy of Sciences, Radzikowskiego 152, 31-342 Krak\'{o}w, Poland}

\begin{abstract}
The Fermi surface can be changed by different external conditions like, e.g., pressure or doping.
It can lead to a change in the Fermi surface topology, called as the Lifshitz transition.
Here, we briefly describe the Lifshitz transitions induced by the external magnetic field in a one-dimensional optical lattice and iron-based superconductors.
We also discuss physical consequences emerging from these transitions.
\end{abstract}

\maketitle

\section{Introduction}

\begin{figure}[!b]
\includegraphics[width=\linewidth,keepaspectratio]{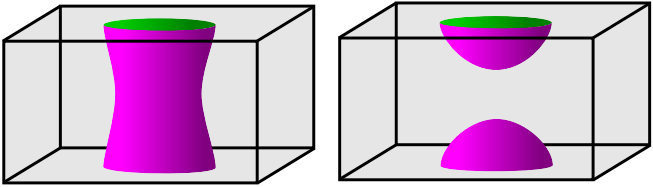}
\caption{A schematic example of the main idea of the {\it Lifshitz transition}.
Here, the Lifshitz transition changes the topology of the Fermi surface from the {\it quasi}-two dimensional one (left) into ''true'' three dimensional one (right).
A black box denotes the first Brillouin zone.
\label{fig.lt}
}
\end{figure}

In some situations specific external conditions can lead to a change of the Fermi surface (FS) topology (Fig.~\ref{fig.lt}).
The main idea of this behavior, called as the {\it Lifshitz transition} (LT), was proposed for the first time by I. M. Lifshitz in his original paper~\cite{lifshitz.60}.
A source of the LT can be, e.g., external pressure or doping.

Since the first description of the LT, this type of behavior has been observed experimentally multiple times in different types of materials, such like, e.g., cuprates~\cite{norman.lin.10,benhabib.sacuto.15,leboeuf.doiron.11}, heavy-fermion compounds~\cite{daou.bergemann.06,harrison.sebastian.07,purcell.graf.09,schlottmann.11,pfau.daou.13,aoki.seyfarth.16}, iron-based superconductors (IBSC)~\cite{khan.duane.14,hodovanets.liu.14,cho.konczykowski.16,liu.lograsso.14,sato.nakayama.09,
nakayma.sato.11,malaeb.shimojima.12,xu.richard.13,liu.kondo.10}, topological insulators~\cite{volovik.17} or graphene~\cite{iorsh.dini.17}.

It should be noted, that in every case the LT is associated with some changes of physical properties of the system.
The influence of the LT on the physical properties will be shortly described in two cases: {\it (i)} Na$_{x}$CoO$_{2}$~\cite{okamoto.nishio.10} and {\it (ii)} doped IBSC from 122 family (i.e. Ba$_{1-x}$K$_{x}$Fe$_{2}$As$_{2}$~\cite{sato.nakayama.09,liu.lograsso.14,khan.duane.14,xu.richard.13,malaeb.shimojima.12,nakayma.sato.11,hodovanets.liu.14,cho.konczykowski.16} or Ba(Fe$_{1-x}$Co$_{x}$)$_{2}$As$_{2}$~\cite{liu.kondo.10}).
In case {\it (i)} we can observe the discontinuous LT~\cite{okamoto.nishio.10}.
For Na$_{x}$CoO$_{2}$ a critical value of the doping $x_{c}$, for which the Fermi level touches the bottom of the band, can be found.
For $x > x_{c}$ a small electron pocket appear.
As a consequence  the dependence of the heat capacity divided by temperature $C/T$ versus $T^{2}$ changes its shape.
A constant Sommerfeld coefficient $\gamma$  for $x < x_{c}$ means that the added electrons occupy the large FS with two-dimensional character, whereas increasing $\gamma$ for $x > x_{c}$ implies that an additional FS emerges.
In case {\it (ii)}, IBSCs from 122 family exhibits a strong dependence of the FS on the doping~\cite{sato.nakayama.09,liu.kondo.10,nakayma.sato.11,malaeb.shimojima.12,xu.richard.13,khan.duane.14,liu.lograsso.14,hodovanets.liu.14,cho.konczykowski.16}.
The angle-resolved photoemission spectroscopy (ARPES) measurements show a change of the FS shape around $M$ point of the first Brillouin zone~\cite{sato.nakayama.09,liu.kondo.10,xu.richard.13,malaeb.shimojima.12,nakayma.sato.11}.
In these compounds the LT induced by doping is observed, e.g., in a low temperature measurement of the Hall coefficient~\cite{liu.kondo.10,liu.lograsso.14} or thermoelectric power~\cite{hodovanets.liu.14}.
Moreover, as consequence  we can observe a change of the superconducting gap from a nodeless one to that of a nodal symmetry~\cite{cho.konczykowski.16}.

In this paper we briefly describe the LT induced by external magnetic field, thus it is called as the {\it magnetic Lifshitz transition} (MLT)~\cite{ptok.kapcia.17,ptok.cichy.17}.
First, we shortly describe the main idea of the MLT (Sect.~\ref{sec.mlt}).
Next, we review two examples of the MLT: {\it (i)} in a case of the one-dimensional periodic chain (Sect.~\ref{sec.mlt1d}) and {\it (ii)} in a case of the two-band model of iron-based superconductors (Sect.~\ref{sec.mlt2d}).
Finally, in Sect.~\ref{sec.sum}, we summarize the results and give a brief outlook.

\section{Magnetic Lifshitz transitions}
\label{sec.mlt}

The MLT in a relatively simply way can be realized in a system, in which the top or bottom of a band is located near the Fermi level.
Then, an external magnetic field $h$, by the Zeeman effect, leads to a shift of energy levels of electrons with opposite spins.
For relatively large $h$  the FS for one spin can be shifted above (below) the Fermi level, what in a case of the electron-like (hole-like) band leads to disappearance of the FS pocket created by this band.
As a consequence of this behavior, similarly like in a case of the standard LT induced, e.g., by doping or pressure, evidences of that transition can be observed in some changes of physical properties of the system.

In literature also other types of the LT induced by magnetic field have been discussed~\cite{purcell.graf.09,schlottmann.11,leboeuf.doiron.11,pfau.daou.13}.
We can mention here a few relevant cases:
{\it (i)} YbRh$_{2}$Si$_{2}$, where the multiple LT induced by a magnetic field has been reported ~\cite{pfau.daou.13}
{\it (ii)} CeIn$_{3}$, where the Fermi surface reconstruction occurs inside the N\'{e}el antiferromagnetic long-range ordered phase~\cite{schlottmann.11,purcell.graf.09}; and
{\it (iii)} YBa$_{2}$Cu$_{3}$O$_{y}$, where the metal-insulator crossover is driven by a Lifshitz transition~\cite{pfau.daou.13}.
However, these issues are beyond the scope of this paper.

\subsection{Magnetic Lifshitz transition in 1D optical lattice}
\label{sec.mlt1d}

The first discussed example of the system, where we can realize the MLT, is the one-dimensional (1D) periodic optical lattice~\cite{ptok.cichy.17,ptok.cichy.17a}.
This system can be described  in the real space by the following Hamiltonian:
\begin{eqnarray}
\label{eq.ham.re.1d} \mathcal{H} &=& \sum_{ i j \sigma } \left( -t \delta_{\langle i,j \rangle} - ( \mu + \sigma h ) \delta_{ij} \right) c_{i\sigma}^{\dagger} c_{j\sigma} \\
\nonumber &+& U \sum_{i} c_{i\uparrow}^{\dagger} c_{i\uparrow} c_{i\downarrow}^{\dagger} c_{i\downarrow} ,
\end{eqnarray}
where $c_{i\sigma}^{\dagger}$ ($c_{i\sigma}$) is the creation (annihilation) operator of the particle with spin $\sigma$ at {\it i}-th site.
Numerically, we express $\sigma$ as $+1$ ($-1$) for the spin parallel (antiparallel) to the external magnetic field $h$ [which corresponds to spin $\uparrow$ ($\downarrow$), respectively].
The first term describes free particles, where $t$ is the hopping integral between the nearest-neighbor sites, $\mu$ is the chemical potential and $h$ is the external magnetic field (in the Zeeman form).
Our main scope of the study are systems in the Pauli limit~\cite{clogston.62,chandrasekhar.62}, in which the orbital effects are sufficiently weaker than the paramagnetic ones~\cite{maki.66}.
On the other hand, this assumption can be realized also in a case when the external magnetic field is directed along the 1D lattice (or in plane the 2D lattice).
The second term describes the on-site pairing interaction ($U < 0$).
This term can be decoupled by mean field approximation, which leads to
\begin{eqnarray}
c_{i\uparrow}^{\dagger} c_{i\uparrow} c_{i\downarrow}^{\dagger} c_{i\downarrow} = \Delta_{i}^{\ast} c_{i\downarrow} c_{i\uparrow} + \Delta_{i} c_{i\uparrow}^{\dagger} c_{i\downarrow}^{\dagger} - | \Delta_{i} |^{2} ,
\end{eqnarray}
where we introduce $\Delta_{i} = \langle c_{i\downarrow} c_{i\uparrow} \rangle$ as a superconducting order parameter (SOP).
Because we assume a general expression for the SOP in the real space, we can take $\Delta_{i} = \Delta_{0} \exp ( i {\bm Q} \cdot {\bm R}_{i} )$, where $\Delta_{0}$ is the amplitude of the SOP, whereas the ${\bm Q}$ is the total momentum of the Cooper pairs.
$\Delta_{0}$ can be treated as an order parameter in the momentum space and the superconducting state exists in the whole system of $\Delta_{0} > 0$ and normal state (NO) occurs otherwise, whereas ${\bm Q}$ describes a type of the superconducting phase \cite{ptok.cichy.17,ptok.cichy.17a}.
If $| {\bm Q} | = 0$ we have NO or BCS phases (with a constant value of the SOP in the real space), whereas otherwise the Fulde--Ferrell--Larkin--Ovchinnikov~\cite{FF,LO} (FFLO) phase is present.
In the momentum space, Hamiltonian~(\ref{eq.ham.re.1d}) can be rewritten in the form:
\begin{eqnarray}
\label{eq.ham.mom.1d}  \mathcal{H} &=& \sum_{{\bm k}\sigma} \mathcal{E}_{{\bm k}\sigma} c_{{\bm k}\sigma}^{\dagger} c_{{\bm k}\sigma} + U \sum_{\bm k} \left(
\Delta_{0}^{\ast} c_{-{\bm k}+{\bm Q}\downarrow} c_{{\bm k}\uparrow} + h.c. \right),
\end{eqnarray}
where $\mathcal{E}_{{\bm k}\sigma} = -2 t \cos ( k_{x} ) - ( \mu + \sigma h )$ is the spin-dependent dispersion relation.
Using the standard Bogoliubov transformation the eigenstates of Hamiltonian~(\ref{eq.ham.mom.1d}) can be found, which allow to calculate the grand canonical potential $\Omega = - k_{B} T \Ln \{ \Tr [ \exp ( - H / k_{B} T ) ] \}$ (details of calculations can be found, e.g., in Ref.~\cite{ptok.cichy.17}).

In the described case, $\Omega$ for fixed physical quantities (i.e., $\mu$, $h$ and $T$) is the function of the $\Delta_{0}$ and ${\bm Q}$, whereas the ground state is defined by the global minimum of $\Omega \equiv \Omega ( \Delta_{0} , {\bm Q} )$.
In a high-magnetic-field regime we expect a discontinuous phase transition~\cite{ptok.15}, thus a correct critical parameters (describing a point of the phase transition) can be found only from  the global minimum of $\Omega$.
It is a consequence of the exact form of the $\Omega$ function~\cite{casalbuoni.nardulli.04}, for which conditions $\partial \Omega / \partial \Delta_{0} = 0$ and $\partial \Omega / \partial | {\bm Q} | = 0$ can give incorrect transition point to the metastable phase, e.g., Sarma phase~\cite{sarma.63}.
Moreover, for a lattice model, where we effectively describe a chain of $N$ sites, we can realize the $N$ different types of the FFLO phase (with $N$ different ${\bm Q}$).
Note that parameters ($\Delta_{0}$ and ${\bm Q}$) describing the ground state of the system strongly depend on the size $N$~\cite{ptok.crivelli.17}.
Thus, it is necessary to adopt a relatively large system for the model.
To accelerate the numerical calculation of the ground state for such a large system, we used the Graphics Processing Unit (GPU) supported CUDA nVidia technology and the procedure described in Ref.~\cite{januszewski.ptok.15}.

\begin{figure}[!b]
\includegraphics[width=\linewidth,keepaspectratio]{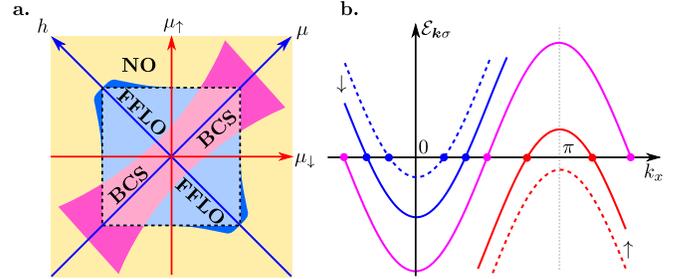}
\caption{
(a) A schematic phase diagram of the system as a function of the effective chemical potentials $\mu_{\sigma} = \mu + \sigma h$ of particles with given spins (red axes) and chemical potential $\mu$ vs. magnetic field $h$ (blue axes) (cf. Ref.~\cite{ptok.cichy.17}).
A black dashed square denotes the bands edges.
Labels BCS, FFLO and NO denote stable phases.
Panel (b) shows the main idea of the magnetic Lifshitz transition in one-dimensional periodic optical lattice.
At half-filling ($\mu = 0$, $n = 1$) and in an absence of the external magnetic field ($h = 0$), we have one two-folded spin-degenerated band with dispersion relation $\mathcal{E}_{{\bm k}\sigma} = -2 t \cos ( k_{x} )$ (pink line).
For $h \neq 0$, the spin-degeneracy is removed (blue and red line correspond to  particles with spins $\uparrow$ and $\downarrow$, respectively).
A dispersion relation is given as $\mathcal{E}_{{\bm k}\sigma} = -2 t \cos ( k_{x} ) - ( \mu + \sigma h )$.
Around half-filling, if the magnetic field is smaller than half-bandwidth ($h < 2 t$), both bands (solid blue and red lines) cross the Fermi level (point on horizontal axis) and two separated Fermi surfaces for electrons with opposite spins can be observed.
If the magnetic field is bigger than a value of the field at the magnetic Lifshitz transition (i.e., $h = 2 t$), only one of the shifted bands (dashed lines) creates the Fermi surface (red points).
\label{fig.1d}
}
\end{figure}

\begin{figure*}[!t]
\includegraphics[width=\linewidth,keepaspectratio]{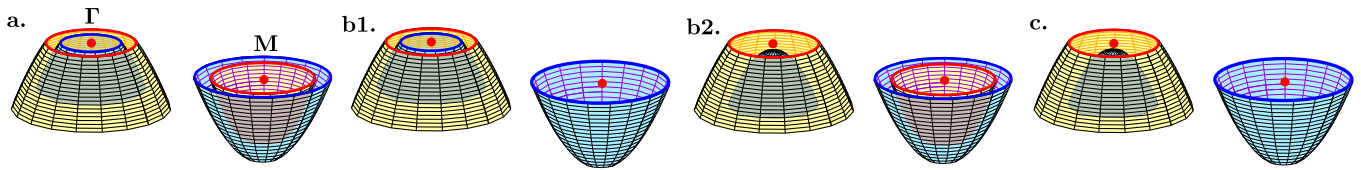}
\caption{
Realizations of the magnetic Lifshitz transition in a case of the iron-based superconductors.
This superconductors is characterized by the band structure with hole-like and electron-like bands around $\Gamma$ and $M$ points of the first Brillouin zone.
It gives the Fermi surface composed from two pockets centered at these points.
(a) A presence of the external magnetic field leads to a shift of energy of electrons with spin up (blue line) and down (red lines).
Thus, four Fermi surfaces exists (solid lines).
Increasing of the magnetic field leads to the situation when one of the Fermi surfaces disappears ((b1) and (b2)).
(c) In an extreme case, for relatively large magnetic fields,  only two Fermi surfaces  are present.
\label{fig.2d}
}
\end{figure*}

A schematic representation of the ground state phase diagram is shown in Fig.~\ref{fig.1d}(a) (cf. with Fig. 4 in Ref.~\cite{ptok.cichy.17}).
We can discuss it in two types of coordinates: {\it (i)} effective chemical potentials of particles with given spins $\mu_{\sigma} = \mu + \sigma h$ (red axes) and/or {\it (ii)} chemical potential $\mu$ vs. magnetic field $h$ (blue axes).
In a case of non-interacting particles, the critical parameters describing the band edges are shown as the dashed black square ($| \mu_{\sigma} | = 2 t$).
Inside this region, the average number of particles $n = 1 / N \sum_{i\sigma} \langle c_{i\sigma}^{\dagger} c_{i\sigma}\rangle$, changing from zero (fully empty bands $\mu_{\sigma} = - 2 t$) to two (fully filled bands $\mu_{\sigma} = 2 t$).
In an absence of the magnetic field ($h = 0$), along the chemical potential $\mu$ coordinate, the BCS phase is stable (pink region).
However, with an increase of the pairing interaction $U$, the region of this phase extends also outside of the dashed square (dark pink region).
This behavior is known as the {\it Leggett condition}~\cite{leggett.80} and describes the BEC--BCS crossover~\cite{micnas.ranninger.90,chen.stajic.05,kujawacichy.micnas.11,cichy.micnas.14,kapcia.14a,kapcia.14b}.
In this regime, relatively large pairing interaction $| U |$ leads to modification of the (non-polarized) quasi-particle spectrum, and consequently to the existence of tightly bound local pairs, even if formally any Fermi surfaces do not exist in the system.
On other hand, if $h > 0$ the FFLO phase can be stable, in which the Cooper pairs have non-zero total momentum (blue region).
This phase, even for weak interaction $| U |$, covers almost the whole range of $\mu$ and $h$ parameters~\cite{koponen.paananen.08,ptok.cichy.17}, what is a result of the ideal Fermi surface nesting in a 1D system.
However, an increase of $U$ leads to leaking of the FFLO phase beyond the boundary for the non-interacting system (dark blue region).
The boundary restricting FFLO phase region  outside the dashed squre  near the half-filling (around $h$ axis), is a condition for MLT emergence and it corresponds to the {\it Leggett condition} in the BEC--BCS crossover case.
Similarly to the previous case, strong $| U |$ modifies the polarized quasi-particle spectrum~\cite{chubukov.eremin.16}, which effectively leads to the situation when one of the spin-type Fermi surface disappears (panel (b))~\cite{ptok.cichy.17}.
Moreover, in this region the FFLO is characterized by $| {\bm Q} | = \pi$~\cite{ptok.cichy.17a}.
This type of the FFLO phase, in which total momentum of the Cooper pair are characterized by the vertex of the first Brillouin zone~\cite{ptok.maska.09}, is called the $\eta$ phase~\cite{kapcia.czart.16,yang.62,ptok.mierzejewski.08,ptok.kapcia.15}.

\subsection{Magnetic Lifshitz transition in iron-based superconductors}
\label{sec.mlt2d}

Now, we will describe the MLT in a case of the IBSCs.
These materials are characterized by the multi-band structure~\cite{kordyuk.12,liu.zao.15}, which is a consequence of layered structure~\cite{stewart.11,liu.zao.15}.
Unfortunately, a source of the superconductivity in these compounds is still unknown~\cite{hirschfeld.korshunov.11}.

Similarly to the one-dimensional system~(Sect.~\ref{fig.1d}), some types of the IBSCs show a tendency to exhibit the FFLO phase~\cite{ptok.15,januszewski.ptok.15,ptok.crivelli.13,zocco.grube.13,cho.yang.17} and the BCS--BEC crossover~\cite{lubashevsky.lahoud.12,kasahara.watashige.14,okazaki.iko.14,kasahara.yamashita.16,rinott.chashka.17}.
The second possibility is a result of the proximity between the  bottom (or top) of the band and the Fermi level, which is found in the Fe(Te,Se) superconductor.

Now we briefly describe how the MLT is realized in the IBSCs.
As a consequence of the layered structure of the IBSCs, the Fermi surface is quasi-two dimensional one composed of hole- and electron-like pockets, around $\Gamma$ and M points of the first Brillouin zone (Fig.~\ref{fig.2d}).
Relatively weak magnetic field $h$ leads to the emergence of four Fermi surfaces (panel (a)), corresponding to given spins.
If $h$ is relatively large (bigger than a distance between the top of band and the Fermi level in a case of the hole-like band and between the bottom of the band and the Fermi level in a case of the electron-like band), we can only observe two of the different spin-type Fermi surfaces (panel (c)).
In the transition region (between (a) and (c) panels), dependently on relation between hole- and electron-like bands widths, we find a range of $h$ in which three Fermi surfaces exist (panels (b1) or (b2)).

A realization of the MLT in the multi-band system can be a source of additional phase transitions inside superconducting phase.
Such a situation has been reported in Ref.~\cite{kasahara.watashige.14}, and observed in the thermal conductivity measurements in a presence of high magnetic field and at low temperature.
The phase transition mentioned above, exists in approximately constant magnetic field independently of temperature, what is in agreement with theoretical calculation~\cite{ptok.kapcia.17}.

\section{Summary}
\label{sec.sum}

The Lifshitz transition can be induced by e.g. doping or pressure.
In some compounds, in which the top or bottom of the band is located near the Fermi level, also an external magnetic field can lead to the Lifshitz transition.
In this paper, we shortly characterized the Lifshitz transitions induced by the  magnetic field and called them as the {\it magnetic Lifshitz transition} (MLT).
We also described two systems, in which the MLT can be realized.

In the first case of one-dimensional optical lattice with attractive interaction, the unconventional pairing can exist in a system for high magnetic fields.
For the strong pairing interaction, in a wide range of magnetic field and doping, one of the spin-type Fermi surface disappears.

In second case of iron-based superconductors the situation is more complicated.
The Fermi surface of these compounds can be characterized by the hole- and electron-like pockets.
If the top or bottom of the bands are near the Fermi level, relatively large magnetic field can lead to the few Lifshitz transitions that are associated with a disappearance of some spin-dependent Fermi surfaces.

\begin{acknowledgments}
The author is thankful to A. Cichy, K. J. Kapcia, A. Kobia\l{}ka, A. M. Ole\'{s}, P. Piekarz and K. Rodr\'{i}guez for very fruitful discussions and comments.
This work was supported by the National Science Centre (NCN, Poland) under grant UMO-2017/25/B/ST3/02586.
\end{acknowledgments}

\bibliography{biblio}

\end{document}